\documentclass[onecolumn,preprintnumbers,superscriptaddress,amsmath,amssymb, bibnotes,nofootinbib]{revtex4}
\usepackage{amsthm}
\usepackage{amsmath}
\usepackage{amssymb}
\usepackage{graphicx}
\usepackage{dsfont}
\usepackage{float}
\usepackage{bm}
\usepackage[usenames,dvipsnames]{color}
\usepackage{mathrsfs}
\usepackage[colorlinks=true,citecolor=blue,urlcolor=black]{hyperref}
\usepackage{float}
\usepackage{adjustbox}
\usepackage{times}
\usepackage{multirow} 
\usepackage{tabularx}

\begin{document}
	
%title
\title{Efficient experimental quantum fingerprinting with wavelength division multiplexing}
\author{Xiaoqing Zhong}
\email{xzhong@physics.utoronto.ca}
\affiliation{Center for Quantum Information and Quantum Control, Dept. of Physics, University of Toronto, Toronto, Ontario, M5S 1A7, Canada}
\author{Feihu Xu}
\affiliation{Hefei National Laboratory for Physical Sciences at the Microscale and Department of Modern Physics, University of Science and Technology of China, Hefei 230026, China}
\author{Hoi-Kwong Lo}
\affiliation{Center for Quantum Information and Quantum Control, Dept. of Physics, University of Toronto, Toronto, Ontario, M5S 1A7, Canada}
\affiliation{Center for Quantum Information and Quantum Control, Dept. of Electrical \& Computer Engineering, University of Toronto, Toronto, Ontario, M5S 3G4, Canada}
\author{Li Qian}
\affiliation{Center for Quantum Information and Quantum Control, Dept. of Electrical \& Computer Engineering, University of Toronto, Toronto, Ontario, M5S 3G4, Canada}

\begin{abstract}
	Quantum communication complexity studies the efficiency of information communication (that is, the minimum amount of communication required to achieve a certain task) using quantum states. One representative example is quantum fingerprinting, in which the simultaneous message passing model is considered and the minimum amount of communication could be exponentially smaller than the classical fingerprinting protocol. Experimental demonstrations based on a practical quantum fingerprinting protocol where coherent states are used to construct the fingerprints, have successfully shown the superiority of quantum fingerprinting. However, as a consequence of using coherent states, the communication time in this coherent quantum fingerprinting (CQF) protocol is increased. Moreover, the minimum amount of information communicated in these experimental demonstrations is largely limited by the dark counts of the single photon detectors. Here, we propose to enhance the performance of the existing CQF protocol through applying wavelength division multiplexing (WDM) and \textit{simultaneous} detection of multiple wavelength channels. We show that the new WDM-CQF protocol can potentially reduce the communication time significantly. More importantly, with the same experimental parameters, the amount of communication is much reduced in the new scheme compared with the original CQF protocol. The more wavelength channels are used, the less communication is required by WDM-CQF protocol. We also perform a proof-of-concept experimental demonstration of the new WDM-CQF protocol with 6 wavelength channels. The experimental results further validate that the new scheme with WDM not only beats the classical protocol, but also reduces the amount of communication required by the original CQF protocol by more than half.
\end{abstract}

\maketitle

\section{Introduction}

	Quantum communication is the study of information transmission tasks that can be facilitated by using quantum mechanical systems~\cite{com_rev}. The power of quantum mechanics enables quantum communication to  perform tasks that could not be accomplished in a classical system. One of the best known examples is quantum cryptography which enables information-theoretically secure communication between two parties that share random keys through quantum key distribution (QKD)~\cite{QC1,QC2,QC3,QKD1,QKD2}. Apart from quantum cryptography, quantum communication complexity (QCC)~\cite{QCC1,QCC2,QCC3,QCC4} is another important example which shows quantum superiority over its classical counterpart - classical communication complexity~\cite{CC1,CC2,CC3,CC4}. In the basic model of communication complexity~\cite{CC3}, Alice and Bob each is given an $n$-bit string $x$ and $y$ respectively. The classical communication complexity exploits the minimum amount of communication necessary among participants, namely the minimum number of bits of communication, such that they could compute a certain function $f(x,y)$ correctly. This exploitation of minimum amount of communication provides a lower bound for many related research area, such as the study of VLSI circuit design, data structure and computer networks~\cite{CC3,CC4}. In the quantum version of communication complexity, the involved participants are allowed to communicate with quantum states instead of classical bits and QCC is then defined as the minimum number of qubits of communication exchanged between Alice and Bob~\cite{QCC4}. It has been proven that, by using quantum superposition or entanglement, many quantum protocols of communication complexity are more efficient, that is, they requiring less communication (fewer qubits) than their classical counterparts~\cite{prove1,prove2,prove3,prove4,prove5,prove6}. 

 	One remarkable protocol in QCC is quantum fingerprinting (QF)~\cite{QF1,QF2} where quantum mechanics can help reducing the communication complexity exponentially compared with the classical case. In the fingerprinting mechanism, the simultaneous message passing model is considered~\cite{CC1}. In this particular model, Alice and Bob have no shared randomness and are not allowed to communicate with each other. But they want to determine whether their inputs $x$ and $y$ are the same or different. In this case, a third party, the referee (Charlie), is involved and will solve this equality problem based on the inputs' fingerprints that Alice and Bob send to her. The communication complexity in this model is defined as the amount of information communicated between Alice (Bob) and Charlie, which is equivalent to the minimum length of the fingerprints. Note that QF protocol is not concerned with communication security. It has been proved that, without any correlations or entanglement shared among the parties, quantum fingerprints require $O(log_{2}n)$ qubits~\cite{QF1}, which are exponentially smaller than the classical case where $O(\sqrt{n})$ bits are required~\cite{bound1,bound2,bound3}. However, to demonstrate the exponential advantage of quantum fingerprinting, one has to create fingerprints consisting of highly entangled-qubit states~\cite{QF1} which are beyond the reach of current technology. Ref.~\cite{Exp1,Exp2,Exp3} have experimentally demonstrated this protocol only with single qubit to show the feasibility of quantum fingerprinting.  In Ref.~\cite{coherent1}, a more practical QF protocol has been proposed and coherent states are used to construct the fingerprint. The minimum amount of communication required in this protocol is proven to be 	
 	\begin{equation}
 		Q=O(\mu log_2n).
 		\label{Q}
 	\end{equation}
 	For simplicity, we call this coherent quantum fingerprinting protocol as CQF protocol in this letter. The total mean photon number of the fingerprint in this coherent quantum fingerprinting (CQF) protocol is $\mu$. Therefore, for CQF protocol with a fixed $\mu$, the minimum amount of communication can still be exponentially smaller than the classical fingerprinting protocol. Ref.~\cite{c-exp1,c-exp2} have successfully demonstrated the proof-of-principle experiment of CQF protocol and prove that less information is communicated in the CQF system compared with the best-known classical protocol~\cite{bound3}. Nonetheless, this CQF protocol uses a number of optical modes that is proportional to the input size $n$, hence the communication time is quadratically increased compared with the classical system~\cite{c-exp1,c-exp2}. In addition, the minimum amount of communication in CQF protocol has a dependence on $\mu$, a value that has a lower limit due to experimental imperfections~\cite{coherent1}, among which the dark counts from the single photo detector (SPD) (used for Charlie's detection) is a dominant factor~\cite{c-exp1}. As indicated in Ref.~\cite{c-exp2} where superconducting nanowire single photon detectors (SNSPD) with {extremely low dark count rates ($<0.1$Hz)} are applied, the performance of the CQF system is significantly improved and can even beat the classical limit\footnote{This classical limit is optimized in Ref.~\cite{c-exp2}. It describes the lower bound of the amount of communication in any classical fingerprinting systems.}. However, such SNSPDs are costly and may well be impractical to implement on a large scale.

	In order to reduce the communication time and improve the performance of CQF protocol, we propose a new protocol utilizing the wavelength division multiplexing (WDM). We call this new protocol WDM-CQF protocol. As a mature technique, WDM has been widely employed in classical communication systems to broaden the communication bandwidth and improve the communication efficiency~\cite{wdm1,wdm2}. It is natural to extend such an advantage into quantum communication systems. A lot of applications of WDM in quantum communication focus on providing shared infrastructure for both classical and quantum communication. There have been few studies using WDM to enlarge the quantum channel capacity~\cite{qq-wdm,coherent2,qq-wdm2}. Especially in Ref.~\cite{coherent2}, coherent state fingerprints are also used to study another QCC protocol - Euclidean problem, which aims at calculating the Euclidean distance of two real vectors of Alice and Bob. The authors similarly propose to employ multiplexing technique to improve the communication efficiency. However, de-multiplexing followed by multiple individual detection systems are always needed in these studies. In this paper, WDM is used to increase the quantum channel capacity \textit{without de-multiplexing}. All the quantum channels share the same detection system. More importantly, by removing de-multiplexing and detecting the signals from different wavelength channels simultaneously, we reduce the amount of communication required in the original CQF protocol. This is because, compared with CQF protocol, for a same input size $n$, our WDM-CQF system requires a lower value of $\mu$ for generating the coherent fingerprints, thus reducing the amount of communication. With large number of wavelength channels, it is in principle possible for our new scheme to beat the classical limit without using SNSPDs. Here, we perform a proof-of-principle experimental demonstration of WDM-CQF protocol with 6 wavelength channels over 40 km fibers. The experimental results validate that, with WDM applied, the new system not only beats the best-known classical protocol, but also reduces the amount of information communicated in the original CQF protocol by more than half.  

 \begin{figure}[b]
	{\includegraphics[width=0.65\linewidth]{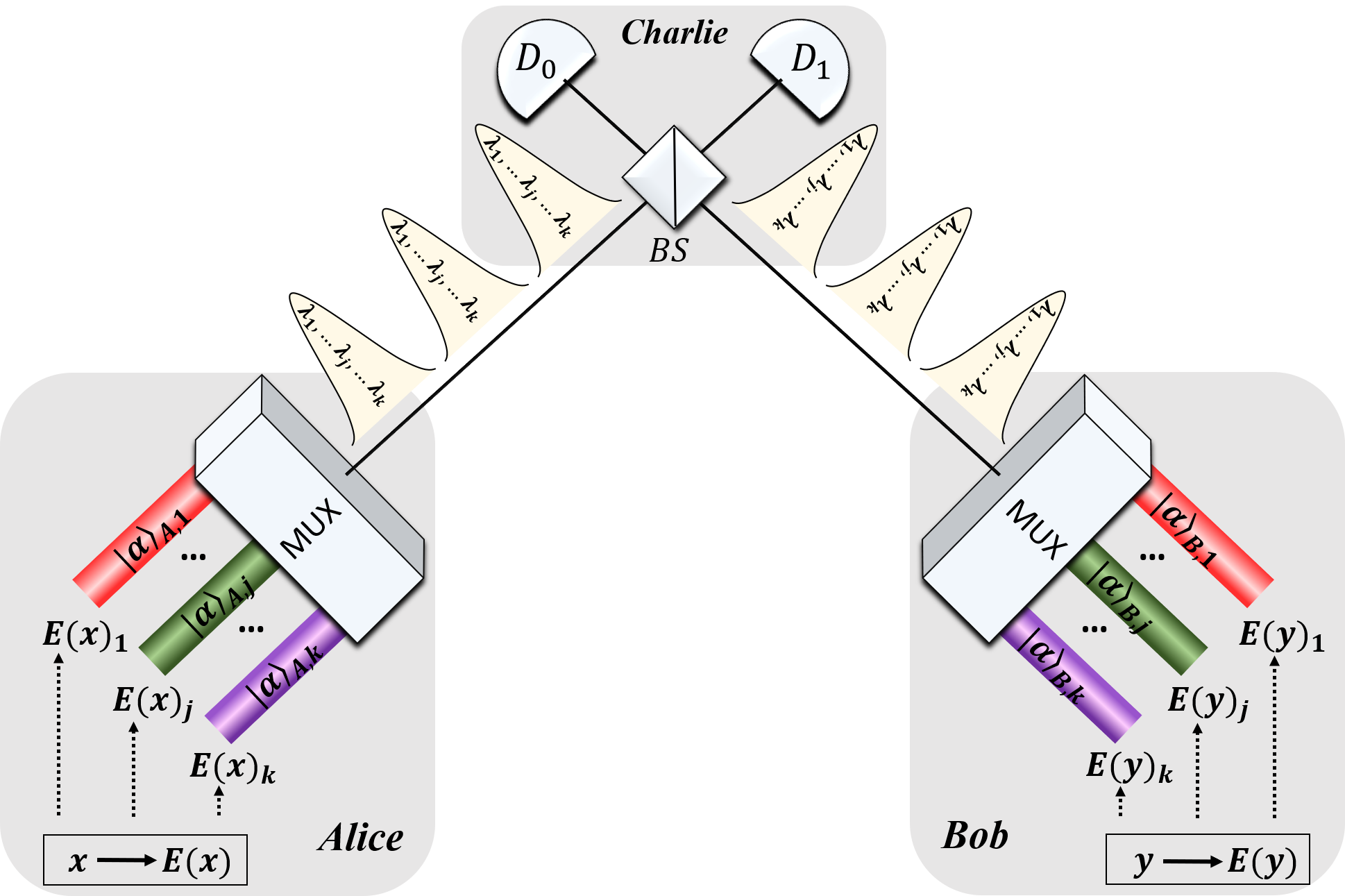}}
	\caption{Theoretical scheme of coherent quantum fingerprinting with wavelength division multiplexing. Alice (Bob) applies error correction code to her (his) input $x (y)$ and obtains $E(x) (E(y))$. Then she (he) divides $E(x) (E(y))$ into k sub-codewords and prepares the corresponding sub-fingerprints ${E_j(x)} ({Ej(y)})$ in k different wavelength channels. The k sub-fingerprints are multiplexed into one single mode fiber through a multiplexer (MUX) and sent to Charlie's beam splitter. On Charlie's station, de-multiplexing is not required. The k pairs of pulses interfere simultaneously and share a pair of single photon detectors $D_0$ and $D_1$. Charlie records the total counts at $D_0$ and $D_1$, based on which Charlie determines whether the inputs are the same or different.} 	
	\label{fig1}
\end{figure}

\section{Method}
	In CQF protocol~\cite{coherent1}, Alice first prepares her coherent fingerprint $\left|\alpha\right\rangle_{A}$ as
 	\begin{equation}\label{cstate}
		\left|\alpha\right\rangle_{A} = \overset{m}{\underset{i=1}{\otimes}}\left|(-1)^{E(x)_i}\frac{\alpha}{\sqrt{m}}\right\rangle_i.
	\end{equation}
	Bob's fingerprint $\left|\alpha\right\rangle_{B}$ has the same expression as Eq.(\ref{cstate}) with changing subscript $A$ into $B$ and changing input $x$ into $y$. The m-bit strings $E(x)$ and $E(y)$ are the codewords of Alice and Bob respectively, obtained by applying error correction code (ECC) to the n-bit input strings $x$ and $y$. The ECC has a code rate $c$ $(=\frac{n}{m}<1)$ and a Hamming distance $\delta$. The use of ECC guarantees that when Alice's and Bob's inputs are different, the minimum number of different bits of $E(x)$ and $E(y)$ is $\delta m$. The codewords contain the information of the original inputs and are encoded into the phase of the coherent states (either 0 or $\pi$ phase is added to the states). As indicated in Eq.~\ref{cstate}, this coherent fingerprint is made up of $m$ coherent states. The mean photon number of each state is $\frac{|\alpha|^2}{m} (=\frac{\mu}{m})$. Then Alice and Bob forward their fingerprints to Charlie's station through two optical channels, where each pair of Alice's and Bob's coherent states interfere with each other and are detected by Charlie's SPDs. Since Alice and Bob each send $m$ coherent states to Charlie, the communication time is proportional to the input size $n/c$. 

	To implement WDM-CQF protocol, Alice and Bob only need to divide their coherent fingerprints into $k$ sub-fingerprints. Each sub-fingerprint consists of $m/k$ coherent states and is described as
 	\begin{equation}\label{sub_state}
 		\left|\alpha\right\rangle_{A,j} = \overset{m/k}{\underset{i=1}{\otimes}}\left|(-1)^{E_j(x)_i}\frac{\alpha}{\sqrt{m}}\right\rangle_i.
 	\end{equation}
	$E_j(x)_i$ is the $i$th bit of the $j$th sub-codeword $E_j(x)$ ($j\in[1,k]$). Fig.(\ref{fig1}) shows the schematic set-up of the WDM-CQF protocol. As shown in Fig.(\ref{fig1}), Alice and Bob assign each sub-fingerprint to a wavelength channel and multiplex the $k$ wavelength channels into a single optical channel. Then  they send their fingerprints to Charlie for detection through the optical channels. In total, $m/k$ wavelength-composite pulses are sent from Alice/Bob to Charlie. On Charlie's side, each pair of the wavelength-composite pulses interfere at the balanced beam splitter (BS) and are measured by two SPDs $D_0$ and $D_1$. Note that, the $k$ pairs of coherent states at different wavelengths in each pulse interfere at the BS independently but simultaneously. Hence the communication time is shortened to $1/k$ times of its original value. We remark that, all the wavelength channels share the same BS and SPDs, thus saving experimental components. Except for adding the additional wavelength channels, the WDM-CQF system is very similar to the original CQF system. In fact, one could treat CQF protocol as a special case of WDM-CQF protocol with single wavelength channel ($k=1$).
 
 	After the measurement, Charlie has to determine whether Alice's and Bob's inputs are the same or not by checking the total counts at $D_0$ and $D_1$. Ideally, if there is any count at $D_1$, the inputs $x$ and $y$ should be different. This is because, if Alice and Bob have the same inputs, their coherent fingerprints are the same and the phase interference of the same states results in clicks only at $D_0$. If the inputs are different, a portion of the interfering states have a $\pi$ phase difference. The photons in these states are registered at $D_1$. However, experimental imperfections, such are dark counts of SPD, would also give clicks in $D_1$ even when the inputs are the same. Here we adopt the decision mechanism introduced in Ref.~\cite{c-exp1}. In Ref.~\cite{c-exp1}, for the equal and different inputs cases, photon counts at detector $D_1$ have the Binomial distributions $B(m,P_E)$ and $B(m,P_D)$ respectively. $P_E$ and $P_D$ are the probabilities of $D_1$ obtaining a click in a single detection window. Based on these distributions, a threshold $C_{1,th}$ is chosen. Charlie then compares the total counts at $D_1$ with $C_{1,th}$. If the total counts is smaller than $C_{1,th}$, Charlie concludes that the inputs are equal. Otherwise, Charlie concludes that the inputs are different. In our WDM-CQF system, since each detection event is independent, the photon counts at $D_1$ also have the Binomial distributions $B(M,P_E)$ and $B(M,P_D)$. Note that $m$ is replaced by $M$, since in total $M$ wavelength-composite pulses are sent from Alice/Bob to Charlie in WDM-CQF protocol. The amplitude of each pulse is 
 	\begin{equation}
 		\label{wdm}
 		\frac{\mu}{m/k}=\frac{\mu}{M}.
	\end{equation}  
 	Ignoring multi-photon contributions, the detection probabilities $P_E$ and $P_D$ are
 	\begin{equation}\label{Pe}
 		P_E=P_{E,signal}+P_{dark}=(1-\nu)(1-e^{-\frac{2\mu\eta}{M}})+P_{dark},
 	\end{equation}
 	\begin{equation}\label{Pd}
 		P_D=P_{D,signal}+P_{dark}=(\delta\nu+(1-\delta)(1-\nu))(1-e^{-\frac{2\mu\eta}{M}})+P_{dark}.
 	\end{equation}
 	For the probability $P_D$, we assume the worst case scenario that the codewords $E(x)$ and $E(y)$ have the minimum distance. $\nu$ is the interference visibility and $\eta$ is the optical channel transmittance. $P_{dark}$ is the dark count probability per detection gate of the SPDs. The error probability for this decision mechanism is 
  	\begin{equation}
  		\label{error}
 		P_{error}=max[P(C_{1,E}>C_{1,th}), P(C_{1,D}<C_{1,th})],
 	\end{equation}
  	and it should be smaller than the tolerable probability $\epsilon$. $C_{1,E}$ and $C_{1,D}$ are the detected total counts at detector $D_1$ for the equal and different inputs cases respectively. As discussed, in Charlie's decision mechanism, the information of which channel carries a photon is not important and only the total counts detected on $D_1$ are valued. Therefore, no de-multiplexing is required on Charlie's side. k pairs of coherent states at different wavelengths interfere simultaneously in each detection window, as indicated in Eq.(\ref{Pe}) and Eq.(\ref{Pd}). For each input size n, the choice of threshold $C_{1,th}$ depends on the total mean photon number $\mu$. As indicated in Eq.(\ref{Pe}) and Eq.(\ref{Pd}), when $\mu$ is so small such that $P_{E/D,signal}<<P_{dark}$, the probabilities $P_E$ and $P_D$ are dominated by $P_{dark}$ and the distributions $B(m/k,P_E)$ and $B(m/k,P_D)$ are fairly close to each other. Consequently, the error probability would be very large. Therefore, large value of $\mu$ is preferred for minimizing the error probability. However, as mentioned before, the amount of communication required by the coherent fingerprint is proportional to the mean photon number $\mu$. So for each input size $n$, one has to balance the two demands of low error probability and small amount of communication. That is to say, one has to find the minimum $\mu$ (and its corresponding threshold $C_{1,th}$) which gives the error probability $P_{error}$ smaller than $\epsilon$. See supplementary for more details~\cite{ref}. 
  
  	\begin{figure}[b]
  		{\includegraphics[width=0.65\linewidth]{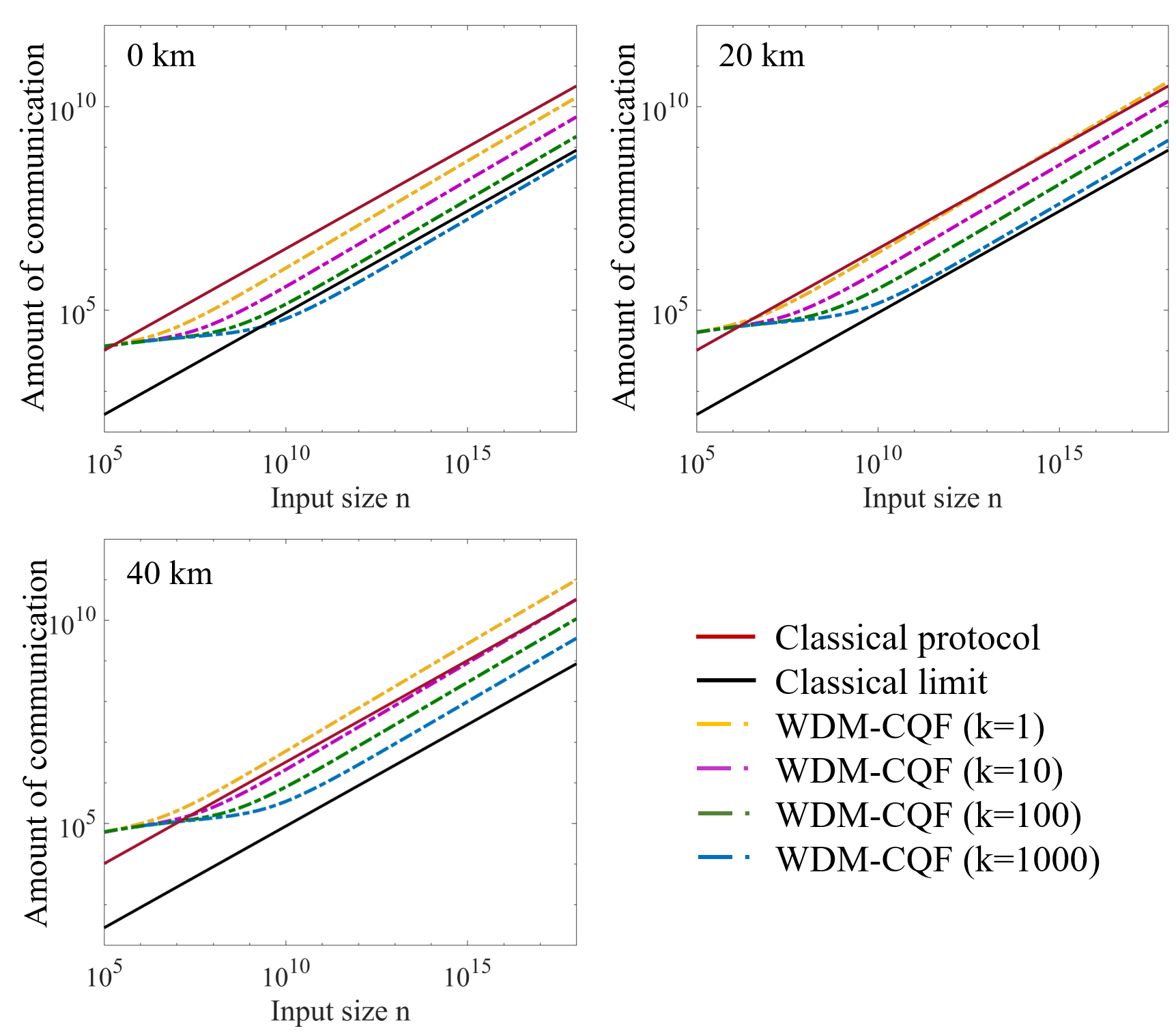}}
  		\caption{ Log-log plot of the simulation results of amount of communication required in different fingerprinting protocols as a function of input size $n$ for three distances, 0km, 20km and 40km. The solid red line represents the best-known classical fingerprinting protocol~\cite{bound3}. The solid black line represents the classical limit introduced in~\cite{c-exp2}. The dash curves are to the simulation results of coherent quantum fingerprinting (CQF) protocol with wavelength division multiplexing (WDM). Different values of $k$ correspond to different number of wavelength channels. When $k=1$, the scheme becomes to the original CQF protocol. In the simulation, we use parameters achievable with single-photon avalanche didoes, with a dark count probability of $1\times10^{-6}$ (per 1 ns detection window) and 20\% detector efficiency. The interference visibility is assumed to be $97\%$. The applied error correction code has a code rate $c=0.24$ of and a Hamming distance $\delta=0.22$. The total mean photon number $\mu$ for each $n$ and $k$ is optimized to fulfill the condition $P_{error}<\epsilon =10^{-5}$. As  indicated in the figure, for all the distances, the CQF system with more wavelength channels required less communication. With 1000 wavelength channels, the system with short distance can even beat the classical limit.} 	
  		\label{qf_simulation}
  	\end{figure}
  
  	It is straightforward to think that the lower the dark count probability $P_{dark}$ is, the smaller $\mu$ can be found. Ref.~\cite{c-exp2} uses SNSPDs with ultralow dark count rate ($0.11$ Hz) and significantly reduces the value of $\mu$, hence can beat the classical limit. However, SNSPDs are much more expensive than the regular SPDs and require extremely low temperature. In fact, instead of lowering $P_{dark}$, our WDM-CQF protocol simply increases the signal probability $P_{E/D,signal}$ due to the simultaneous interference of k pairs of coherent states. Therefore, compared with CQF protocol with single wavelength channel ($k=1$), the new WDM-CQF ($k>1$) not only reduces the communication time, but also requires less $\mu$ to achieve the expected error probability, thus reducing the amount of communication.
  
  	Fig.(\ref{qf_simulation}) shows the amount of communication between Alice/Bob and Charlie over 0 km, 20 km and 40 km fibers in different fingerprinting protocols as a function of the input size $n$. In this log-log plot, practical experimental parameters are considered. The dark count probability $P_{dark}$ is fixed to be $10^{-6}/ns$ and 20\% of detector efficiency is also taken into consideration. The tolerable error probability is chosen to be $\epsilon=10^{-5}$. The input size $n$ varies from $10^5$ to $10^{18}$. (Details about the simulation can be found in Supplementary~\cite{ref}.) As shown in Fig.(\ref{qf_simulation}), when the distance between Alice/Bob and Charlie is 0 km, all the WDM-CQF protocols requires less communication than the best-known classical fingerprinting protocol. As $k$ gets larger, the advantage of WDM-CQF protocol is more evident. When $k=1000$ wavelength channels are applied, indicated by the blue dash line,  the coherent fingerprinting system can even beat the classical limit. When the distance increases, more photons are needed to compensate the channel loss. Hence the amount of communication in the coherent fingerprinting system increases with the channel distance. As depicted in Fig.(\ref{qf_simulation}), the superiority of the original CQF protocol ($k=1$) over classical fingerprinting protocol dramatically diminishes when the distance is 20 km and totally vanishes when the distance is 40 km. While for the new WDM-CQF protocol with $k>1$, the advantage of requiring less communication than the classical protocol is still significant with long distance, especially for large number of wavelength channels. We note that the implementation of WDM-CQF with 1000 channels is challenging though over-1000-channel ultra dense WDM does exist~\cite{DWDM1,DWDM2}. However, the WDM-CQF scheme with 100 channels is promising since there have been many reports of classical transmission experiments with WDM over more than 100 channels~\cite{DWDM3,DWDM4}. Even with 100 channels, our WDM-CQF system can still produce a remarkable advantage over both the best-know classical fingerprinting protocol and the original CQF protocol.  
  	
  	\begin{figure}[b]
  		{\includegraphics[width= 0.65\linewidth]{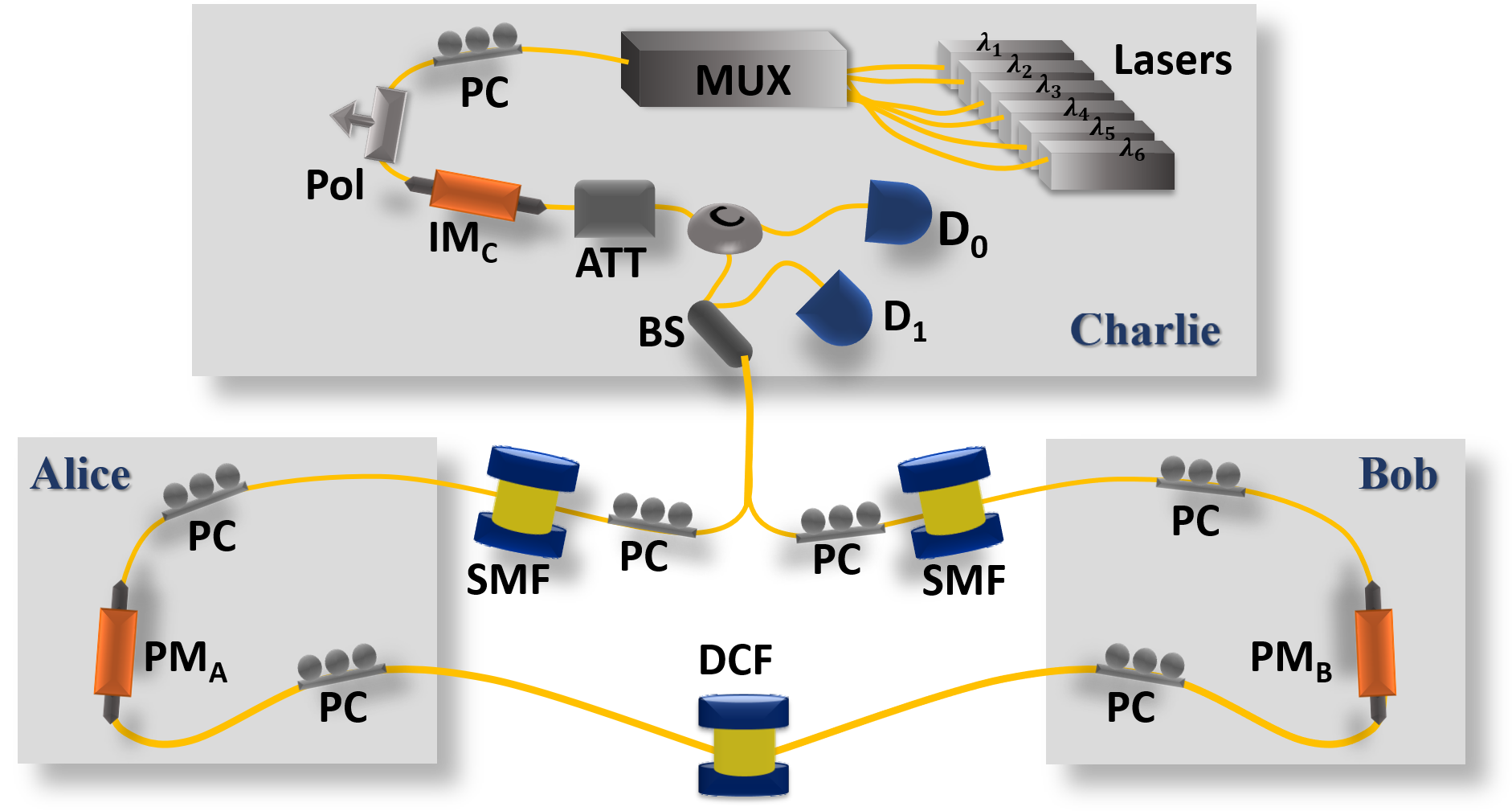}}
  		\caption{ Schematic experimental set-up of coherent quantum fingerprinting with wavelength division multiplexing. Six continuous wave lasers are located on Charlie's side with wavelength ranging from $1542.9 nm$ to $1554.9 nm$, equally spaced by $\delta\lambda=2.4 nm$. Photons coming out the lasers are multiplexed through a multiplexer (Mux) into a single mode fiber and pass through a polarizer (Pol). An intensity modulator (IM) and an optical variable attenuator are used to created weak coherent pulses. The pulses then enter the loop through a circulator (C) and a beam splitter (BS) and travel to Alice/Bob through 20 km single mode fibers (SMF). Alice and Bob are separated by another 6.9 km of compensation dispersion fibers (DCF). On Alice's (Bob's) station, the phase modulator (PM) is on only when the clockwise (counter-clockwise) traveling pulses arrive and the phase information is added to the pulses accordingly. After the phase modulation, the clockwise and counter-clockwise traveling pulses go back to Charlie and interfere with each other at Charlie's beam splitter. The results are recorded by two single photon detectors $D_{0}$ and $D_1$. Polarization controllers (PC) are designed for the polarization alignment for the beams in 6 wavelength channels.} 	
  		\label{fig3}
  	\end{figure}
 
 \section{Experiment}
\begin{figure}[b]
 		{\includegraphics[width= 0.75\linewidth]{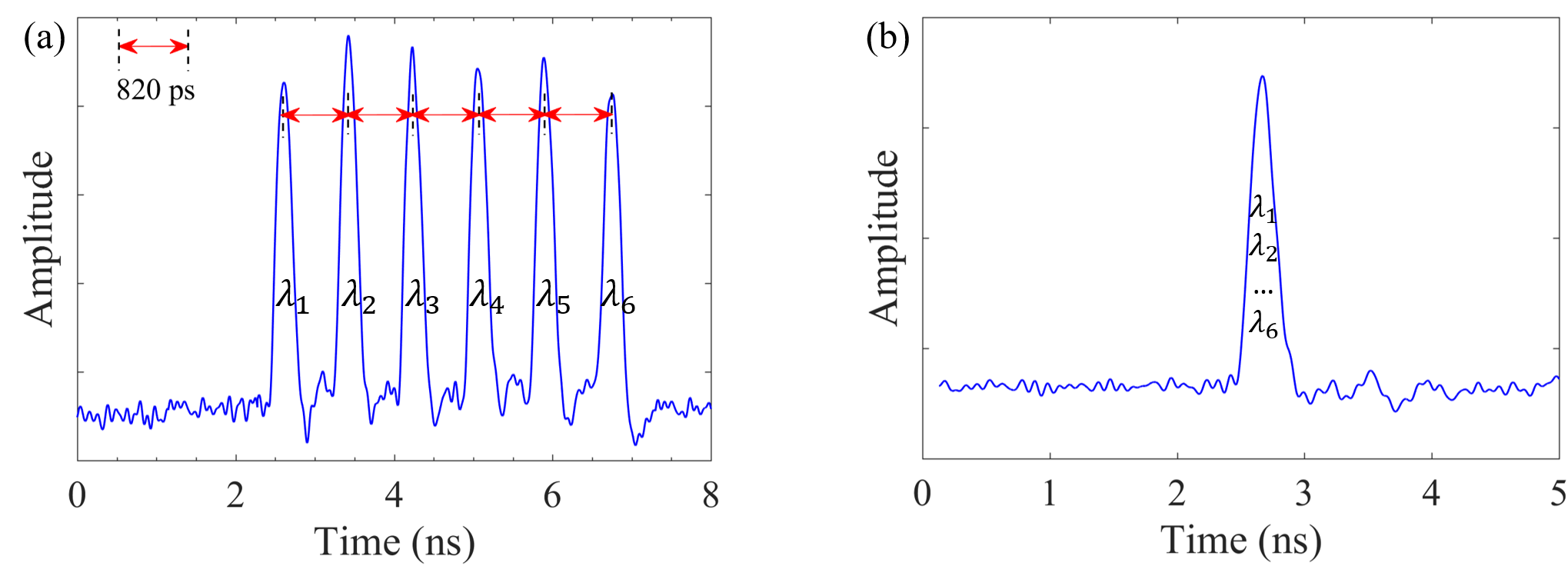}}
 		\caption{Arrival times of different wavelength components. The wavelength components traveling in counter-clockwise direction are (a) temporally separated at Bob's station for ease of individual modulation, while they are (b) combined into the same time slot at Charlie for detection. The components traveling in clockwise direction have the same time distribution since Alice's station and Bob's station are symmetric. Note, one should ignore the slight amplitude difference between the channels as this plot is taken prior to amplitude fine adjustments.}
 		\label{dis_fig}
 	\end{figure}

 	In this section, we show a proof-of-concept experimental demonstration of our WDM-CQF protocol. Six wavelength channels are used and a two-way quantum communication system consisting of a Sagnac interferometor is employed. This system configuration is similar to that of a twin-field QKD system~\cite{tf-qkd}. The Sagnac arrangement is chosen to provide a phase reference between Alice and Bob and to stabilize the phase fluctuation along the optical channel. The schematic experimental set-up is shown in Fig.(\ref{fig3}). On Charlie's station, the continuous waves (cw) coming out of 6 laser modules (PRO 800, wavelength $\lambda \in \left\lbrace 1542.9 nm, 1545.3 nm, 1547.7 nm, 1550.1 nm, 1552.5 nm, 1554.9 nm\right\rbrace$) are multiplexed into a single mode fiber (SMF) through a multiplexer (Jobin Yvon-Spex, Stimax WDM, 100 GHz) and are forwarded to Charlie's intensity modulator (IM$_C$) through a polarizer. IM$_C$ is used together with an optical attenuator (Att$_C$) to create weak wavelength-composite pulses (500 ps pulse width) at a repetition rate of 50 MHz. Then Charlie sends the pulses to Alice and Bob through an optical circulator and a 50:50 fiber-based beam splitter (BS). After passing through the BS, the pulses split into Clockwise and counter-clockwise traveling beams and travel  through a 20-km single mode fiber spool SMF$_B$ or SMF$_A$ respectively. When the clockwise (counter-clockwise) traveling pulses arrive at Bob's (Alice's) station, the fiber based phase modulator PM$_B$ (PM$_A$) is turned off and no information is encoded into the pulses. This guarantees that Alice and Bob have no direct communication. Then the clockwise (counter-clockwise) traveling pulses go through 6.9 km dispersion compensation fibers (DCF) before arriving at Alice's (Bob's) station. Note that this 6.9 km DCF is designed for temporal dispersion. In our system, each pulse created by Charlie has 6 wavelength components. To modulate the phase of each component individually, we use the natural property of fiber, that is chromatic dispersion, to separate these wavelength components in time.  The dispersion parameters (around 1550 nm) of the SMF and DCF in our set-up are $D_{SMF}=17 ps/(nm\cdot km)$ and $D_{DCF}=-99 ps/(nm\cdot km))$ respectively. Given that the wavelength difference between two adjacent modes is $\delta\lambda=2.4nm$, we can estimate the time difference of arrival $\delta T_{A/B}$ at Alice's/Bob's station between the adjacent wavelength components by
	\begin{equation}
 		\delta T_{A/B}= |D_{SMF}\times \delta\lambda\times l_{SMF_{B/A}}+D_{DCF}\times \delta\lambda\times l_{DCF}|.
 		\label{dis}
 	\end{equation}
 	$l_{SMF}$ and $l_{DCF}$ are the lengths of the single mode fibers and dispersion compensation fibers respectively. Fig.~(\ref{dis_fig}a) shows the different arrival times of the 6 wavelength components at Bob's station after traveling through 20 km SMF$_A$ and 6.9 km DCF. As indicated, $\delta T_{A/B}$ in our experiment is around 820 ps, which enables Alice (Bob) to modulate the phases of the 6 wavelength components sequentially by using a 800 ps phase modulation window for each component. After the phase modulation, Alice (Bob) forwards the pulses to Charlie's BS through another 20 km fiber spool SMF$_A$ (SMF$_B$). The length of the DCF is designed to ensure that the, 6 wavelength components overlap with each other in time and become a single wavelength-composite pulse at Charlie's BS. The time difference of arrival at Charlie's station can be estimated by
  	\begin{equation}
 		\delta T_{C}= |D_{SMF}\times \delta\lambda\times (l_{SMF_{A}}+l_{SMF_{B}})+D_{DCF}\times \delta\lambda\times l_{DCF}|,
 		\label{dis1}
 	\end{equation} which is around 0 ps. As shown in Fig.~(\ref{dis_fig}b), after traveling through the whole loop, the six wavelength components arrive at Charlie's BS at the time and overlap with each other.
 	
 	The clockwise and counter-clockwise traveling pulses interfere with each other at Charlie's BS and are detected by two single photon detectors (SPD) $D_0$ and $D_1$. We emphasis that de-multiplexing is not needed on Charlie's station. As mentioned before, the wavelength information of the detected photon is not important. Only the total counts at $D_1$ determines Charlie's output. Therefore, the 6 pairs of coherent states at different wavelengths share the same BS and detectors. \footnote{Here we ignore any cross talk between the adjacent channels and assume that the interference of pluses in each wavelength channel is independent.} The SPDs are commercial avalanche photodiodes (ID220) with an efficiency of 20\% and a dark count rate of 1 kHz. After the measurement, Charlie counts the total number of click events in detector $D_1$ only and compares it with a predetermined threshold value $C_{1,th}$. If the number is smaller than the threshold $C_{1,th}$, he announces that the inputs of Alice and Bob are equal. Otherwise, he concludes that the inputs are different.
 
  	The most challenging problem in our experiment is the wavelength-dependent polarization mode dispersion of long optical fibers~\cite{pol1,pol2,pol3}. Due to the birefringence of optical fiber, the polarization state varies along the fiber. To guarantee the high interference visibility, the polarization of the interfering pulses should be aligned with each other. This alignment could be easily accomplished with one polarization controller (PC) if only one wavelength channel is applied. However, the variation of polarization strongly depends on wavelength, especially for long fibers. Therefore, when multiple wavelength channels are used, the polarization states at different wavelengths evolve differently, making the polarization alignment difficult. As a result, the interference visibility would be affected significantly. To solve this issue, we utilize the principal state of polarization (PSP)~\cite{pol2}. For a fiber system, there are always two orthogonal PSPs, the polarization evolution of which does not depend on the wavelength to the first order\footnote{Note that we ignore the high order polarization mode dispersion since that the wavelength range in our experiment is only 12 nm.}. That is to say, if the input polarization states of the different wavelength components are the same and aligned to the input psp of the optical fibers, the output polarization states should also be the same. In our set-up, there are three long fiber spools and two polarizers (integrated with the phase modulators) used in the Sagnac loop. Therefore, 6 polarization controllers are inserted into the Sagnac loop for the alignment, as shown in Fig.(\ref{fig3}). One PC at one end of a fiber spool is designed to align the input polarization state to the input PSP; the other PC at the other end is used to align the polarization state to the output PSP of the fibers. With such alignment scheme, we are able to maintain our interference visibility to be 97\% over 12 nm bandwidth.
  
  	Another challenge in our implementation is the calibration of the fiber length. First of all, as indicated in Eq.(\ref{dis}), the lengths of SMF and DCF determine the time difference of arrival $\delta T$ among different wavelength components. On one hand, we have to ensure that on Alice's and Bob's stations, $\delta T_{A/B}$ is large enough such that Alice and Bob can modulate the different wavelength components separately; on the other hand, when the pulses travel back to Charlie's station, $\delta T_C$ should be 0 ps. Therefore, the fiber lengths of SMF$_A$, SMF$_B$ and DCF are carefully calibrated to fulfill these two conditions. Additionally, it is also crucial to ensure that the clockwise and counter-clockwise traveling pulses should never 'collide' at Alice's and Bob's phase modulators. This is because when Alice (Bob) should only modulate the clockwise (counter-clockwise) traveling pulses. To avoid the pulse collision, small segments of fibers can be added or deleted on Alice's and Bob's station. Meanwhile, all the phase modulators and intensity modulator are driven and synchronized by a high speed arbitrary waveform generator (AWG, Keysight M8195A). The delays of Alice’s and Bob’s phase modulation signals are well adjusted to ensure that the modulation signals only act on the intended pulses.

\section{Results}

	The experiment was run over seven different values of the input size $n$, ranging from $1.4\times10^6$ to $1.1\times10^9$. For each input size $n$, we tested both the case where the inputs are the same ($x=y$) and the case where the inputs have 1 bit difference ($x\neq y$, $E(x)$ and $E(y)$ have ($\delta m$)-bit difference). The code rate and Hamming distance are $c=0.24$ and $\delta=0.22$. The total mean photon numbers sent out by Alice ($\mu_A$) and Bob ($\mu_B$) are listed in Table (\ref{tab}). The reason why Alice and Bob have different $\mu$ is that the channel loss between Alice and Charlie is slightly different from the loss between Bob and Charlie. Based on the average photon numbers reported, we can determine the threshold value of total counts $C_{1,th}$ at detector $D_1$ as well as its corresponding error probability $P_{error}$. Note that total mean photon numbers in this implementation are close to but not exactly the optimal value. Therefore, the error probabilities for some cases are larger than $\epsilon=10^{-5}$. Nevertheless,the largest $P_{error}$ is $2.7\times 10^{-5}$ which is tolerable~\cite{c-exp1}. The total counts recorded by detector $D_1$ for the equal inputs case ($C_{1,E}$) and the different inputs case ($C_{1,D}$) are also listed in Table (\ref{tab}). For all the seven different input sizes, Charlie could successfully differentiate between the equal inputs and different inputs by comparing the total counts at $D_1$ with the threshold $C_{1,th}$. $Q$ is the amount of communication between Alice/Bob and Charlie in our experiment. To show the advantage of our WDM-CQF protocol, we calculated the ratio $\gamma_C= 32\sqrt{n}/Q$ ($32\sqrt{n}$ is the minimum amount of communication required in the best-known classical fingerprinting protocol~\cite{bound3}), as well as the ratio $\gamma_Q = (the~minimum~amount~of~communication~in~CQF~protocol)/Q$. As shown in Table (\ref{tab}), for all the tested input sizes, $\gamma_C$ and $\gamma_Q$ are always larger than 1, indicating that our WDM-CQF protocol not only requires less communication than the best-known classical protocol, but also beats the original CQF protocol. For large input size, our implementation even reduces more than half of amount of communication in CQF protocol. 

	\begin{table}[t]
	\begin{tabular}{ccccccccccc}
		\hline
		\hline
		{\boldmath $n$}&{ \boldmath $M$} & {\boldmath $\mu_A$} & {\boldmath $\mu_B$} & {\boldmath $C_{1,E}$} & {\boldmath $C_{1,D}$} & {\boldmath $C_{1,th}$}& {\boldmath $p_{error}$} & {\boldmath Q} & {\boldmath $\gamma_C$} & {{\boldmath $\gamma_Q$}} 
		\\
		\hline
		$1.44\times 10^6$& $1.0\times 10^6$ & $1282\pm39$ & $1479\pm45$  & $2.7\pm0.1$  & $34.3\pm0.4$ & $15$ & \parbox[t]{1.6cm}{$(2.7\pm0.8)$\\$\times10^{-5}$} & $37321\pm998$ & $1.03\pm0.03$ & $1.26\pm0.03$
		\\ 
		$2.16\times 10^6$ & $1.5\times 10^6$ & $1425\pm11$ & $1644\pm13$  & $3.0\pm0.2$  & $38.4\pm0.4$ & $16$ & \parbox[t]{1.6cm}{$(1.1\pm0.1)$\\$\times10^{-5}$} & $43792\pm307$ & $1.10\pm0.01$ & $1.22\pm 0.01$
		\\
		$3.60\times 10^7$ & $2.5\times 10^7$ & $2724\pm76$ & $3143\pm88$  & $25\pm2$  & $96\pm2$ & $57$ &\parbox[t]{1.6cm}{$(3.2\pm1.7)$\\$\times10^{-6}$} & $100176\pm2567$ & $1.92\pm0.05$ & $1.64\pm0.04$
		\\
		$7.19\times 10^7$& $5.0\times 10^7$ & $3150\pm105$ & $3635\pm121$  & $53\pm7$  & $130\pm10$ & $87$ & \parbox[t]{1.6cm}{$(1.1\pm0.6)$\\$\times10^{-5}$} & $121232\pm3690$ & $2.24\pm0.07$ & $1.90\pm0.06$
		\\
		$1.44\times 10^8$& $1.0\times 10^8$ & $4050\pm256$ & $4673\pm296$  & $95\pm4$  & $202\pm23$ & $145$ & \parbox[t]{1.6cm}{$(1.6\pm1.0)$\\$\times10^{-5}$} & $161422\pm9406$ & $2.4\pm0.1$ & $2.0\pm 0.1$
		\\
		$3.60\times 10^8$& $2.5\times 10^8$ & $6051\pm469$ & $6982\pm541$  & $251\pm14$  & $395\pm24$ & $309$ & \parbox[t]{1.6cm}{$(1.2\pm0.9)$\\$\times10^{-5}$}  & $250871\pm17973$ & $2.4\pm0.2$ & $2.0\pm0.1$
		\\
		$1.08\times 10^9$& $7.5\times 10^8$ & $9722\pm642$ & $11218\pm741$  & $729\pm41$  & $958\pm43$ & $815$ & \parbox[t]{1.6cm}{$(9.4\pm5.5)$\\$\times10^{-6}$} & $423574\pm14812$ & $2.5\pm0.1$ & $2.15\pm 0.08$
		\\
		\hline
		\hline
	\end{tabular}
		\caption{\label{tab} List of experimental parameters and experimental results. $n$: size of the input string. $M$: total number of wavelength-composite pulses sent from Alice/Bob to Charlie. $\mu_A$ and $\mu_B$: total mean photon numbers of the pulses sent from Alice and Bob; $C_{1,E}$: total counts recorded by detector $D_1$ when Alice and Bob have same inputs; $C_{1,D}$: total counts recorded by detector $D_1$ when Alice and Bob have different inputs; $C_{1,th}$: threshold value of the total counts at detector $D_1$; $Q$: amount of communication in our experiment; $\gamma_C$: ratio of the amount of communication in the best-known classical fingerprinting protocol~\cite{bound3} to Q ($32\sqrt{n}/Q$); $\gamma_Q$: ratio of the amount of communication in the original coherent fingerprinting protocol~\cite{coherent1} to Q; $P_{error}$: error probability.}
	\end{table}

	\begin{figure}[t]
		{\includegraphics[width=0.65\linewidth]{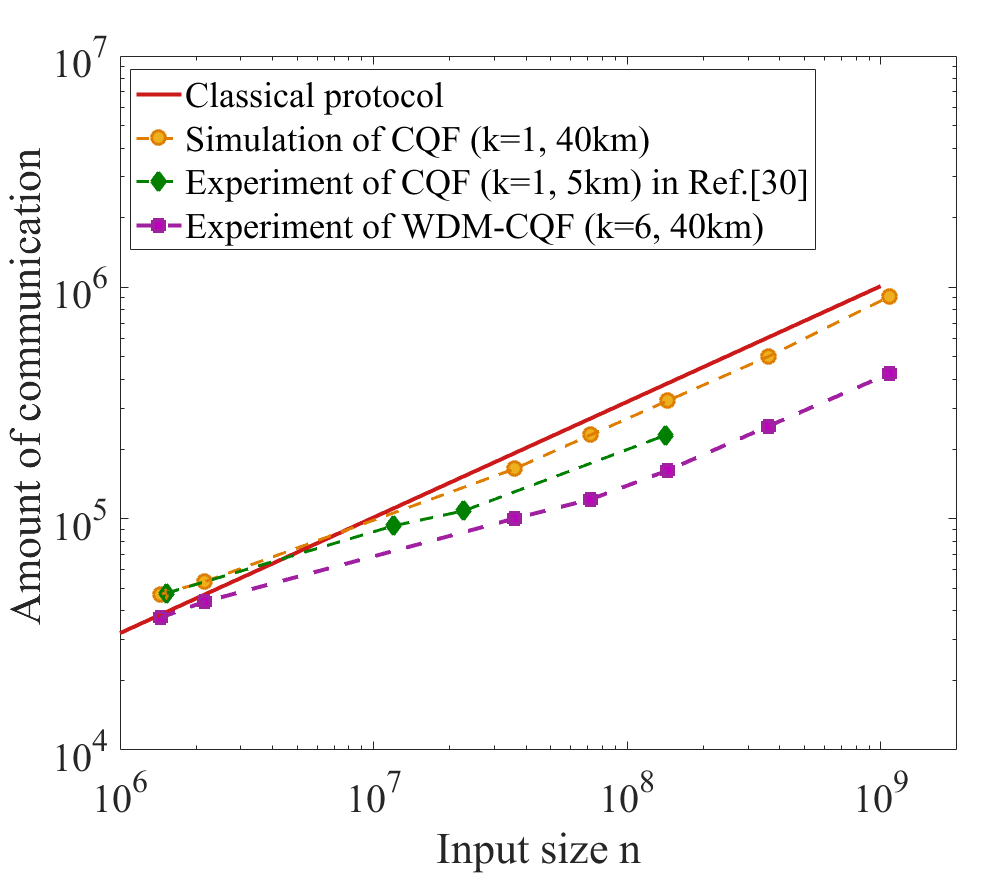}}
		\caption{Log-log plot of the amount of communication between Alice/Bob and Charlie in different fingerprinting protocols, as a function of input size $n$. The solid red curve represents the best-known classical fingerprinting protocol~\cite{bound3}. The purple squares are the amount of communication in our demonstration of coherent quantum fingerprinting (CQF) with wavelength division multiplexing (WDM). Six wavelength channels are used. Except the 6.9 km DCF, the overall distance between Alice and Bob is about 40 km. The orange circles corresponds to the amount of communication in the original CQF system ($k=1$) under the same experimental parameters. It is clear that less information is communicated in our experiment than that in both the classical fingerprinting protocol and the original CQF protocol. We also plot out the amount of communication in another CQF experiment with single wavelength channel~\cite{c-exp1} (green diamonds) for further comparison. Ref.~\cite{c-exp1} uses the same single photon detectors as ours but has much shorter distance (only about 5km). Again, our WDM-CQF system outperforms the original CQF system.} 	
		\label{result}
	\end{figure}

	The experimental results are also illustrated in Fig.(\ref{result}), which is a log-log plot of the amount of communication in different fingerprinting protocols as a function of input size $n$. The solid red line represents the amount of communication required in the best-known classical fingerprinting protocol~\cite{bound3}. Our experimental results are represented by purple squares. The orange circles correspond to the simulation of the original CQF system with the same experimental parameters. Fig.(\ref{result}) clearly shows that our new WDM-CQF system can beat the best-known classical fingerprinting protocol. More importantly, even with only 6 wavelength channels, the new scheme still significantly reduces the amount of communication in the original CQF protocol. For input size $1.44\times 10^6$ and $2.16\times 10^6$, the amount of communication in the CQF system with single wavelength is even higher than the classical system. In our experiment, the amount of communication is always less than the best-known classical protocol. Especially for large input size, the advantage of using WDM is remarkable. For further comparison, we plot the experimental results reported in Ref.\cite{c-exp1} which uses the same SPDs (ID220) to demonstrate the original CQF protocol with single wavelength channel.\footnote{We do not compare with the experimental results in Ref.~\cite{c-exp2}, since the dark count rate of the SNSPD in Ref.~\cite{c-exp2} is four orders of magnitude lower than the dark count rate of the SPD used in our experiment. The main purpose of our experiment is to show that even with regular SPDs, applying WDM can improve the performance of the CQF system.} Note that the total distance implemented in Ref.~\cite{c-exp1} is only about 5 km, which is much shorter than the 40 km total distance in our implementation. Yet, the amount of information communicated in Ref.\cite{c-exp1} is much higher than our experimental results. This comparison further validates the fact that applying WDM can remarkably improve the performance of the original CQF protocol. The new WDM-CQF system is more robust to experimental imperfections (such as dark counts and channel losses). 
 
	Ideally, through applying 6 wavelength channels, the communication time can also be decreased by a factor of 6. However, considering that the 6 wavelength components are phase modulated one by one on Alice's and Bob's stations, our implementation does not strictly shorten the communication time. As a proof-of-concept demonstration, our experiment mainly proves that applying WDM to the CQF system can significantly reduce the amount of communication. Hence, we utilize the inherent chromatic dispersion of single mode fibers to simplify the phase modulation process on Alice's and Bob's stations. To strictly show that WDM-CQF protocol also requires shorter communication time, one can use spatial dispersion rather than temporal dispersion for phase modulation.

	The main limitation of our experiment is the tolerable wavelength channels of our system. In order to avoid the pulse collision during the phase modulation process, the span of different wavelength components on Alice's and Bob's stations should be at most half of the pulse repetition period, that is 10 ns in our system. Given the 800 ps modulation time for each channel and a channel spacing of 2.4 nm, at most 12 wavelength channels with a bandwidth around 26 nm can be applied to our system. One can change the corresponding experimental parameters (such as repetition rate, modulation window and $\delta\lambda$) to increase the tolerable channel numbers. More importantly, through using spatial dispersion to replace temporal dispersion, the above limitation can be removed. When the number of wavelength channel is increased, the current polarization alignment method may not work due to the increased bandwidth. This issue might be solved by inserting a polarizer on each arm of Charlie's BS. The intensities of the pulses in different wavelength channels should also be adjusted by Alice and Bob to compensate the different losses. In our implementation, since the intensity ($\mu/m$) of each wavelength component is very low and channel spacing is not too narrow, we ignore the possible cross talk~\cite{crs1,crs2} effect in our WDM system. Further study about cross talk effect may be necessary if ultra dense WDM (with very small channel spacing $\delta\lambda$) is used. 

	In summary, we propose a new WDM coherent quantum fingerprinting scheme to improve the performance of the existing coherent fingerprinting protocol. We show that by using WDM, our scheme can reduce the communication time of the original CQF protocol.  More importantly, it decreases the amount of communication. Through applying 1000 wavelength channels, the new WDM-CQF scheme with practical experimental conditions can even beat the limit of classical fingerprinting without using SNSPDs. We have performed a proof-of-concept experimental demonstration of the WDM-CQF protocol with 6 wavelength channels. The experimental results clearly show that the new WDM-CQF scheme is robust and significantly outperforms both the classical and coherent fingerprinting protocols. Our practical and economical demonstration of quantum fingerprinting further validates the superiority of quantum communication complexity over its classical counterpart and shows the feasibility of real application. 
	
	We remark that the way we apply WDM to a quantum communication system is unique. De-multiplexing on detection side is removed and signals at different wavelengths are measured simultaneously by a single detecting system. It would be interesting to expand our method to other quantum communication protocols. In fact, in the coherent quantum fingerprinting system, the measurement on Charlie's side is equivalent to a swap test, which has been applied in many other quantum communication protocols, such as quantum digital signature~\cite{qds}. Our study introduces a new method of using WDM to do such a test and shows the feasibility of applying WDM to the other protocols. Last but not least, in our implementation (and also in Ref.~\cite{c-exp1,c-exp2}), a two-way quantum communication system is used to ensure that Alice and Bob have the matched global phase. In this case, Alice's station and Bob's station are actually physically connected. To remove this connection and to enable Alice and Bob independently prepare their fingerprints, one could also employ the method in Ref.~\cite{qf3}, where quantum fingerprinting based on higher order interference is proposed and phase reference is not needed. An interesting question for future study could be whether we can still apply WDM to this method to further improve the communication efficiency. It would also be interesting to explore the possibility of using other degrees of freedom to increase the quantum channel capacity and make quantum communication more efficient.

\section{Acknowledgment}

	We thank Shihan Sajeed, Olinka Bedroya and Wenyuan Wang for their insightful discussion and suggestions. We also thank funding from NSERC, CFI, ORF, MITACS, US ONR, Royal Bank of Canada and Huawei Technologies Canada, Inc.

\end{document}